\documentclass[conference]{IEEEtran}
\IEEEoverridecommandlockouts

\usepackage{tikz}
\usepackage{textcomp}
\usepackage{hyperref}

\newcommand\copyrighttext{{
  \footnotesize \textcopyright 2021 IEEE. Personal use of this material is permitted.
  Permission from IEEE must be obtained for all other uses, in any current or future
  media, including reprinting/republishing this material for advertising or promotional
  purposes, creating new collective works, for resale or redistribution to servers or
  lists, or reuse of any copyrighted component of this work in other works.}}
\newcommand\copyrightnotice{%
\begin{tikzpicture}[remember picture,overlay]
\node[anchor=south,yshift=10pt] at (current page.south)
{\fbox
{\parbox
{\dimexpr\textwidth-\fboxsep-\fboxrule\relax}
{\copyrighttext}
}
};
\end{tikzpicture}%
}

\begin{document}

\title{Versatile and concurrent FPGA-based architecture for practical quantum communication systems}
 \author{Andrea Stanco$^{1,*}$, 
Francesco B. L. Santagiustina$^{1,2}$, 
Luca Calderaro$^1$ 
Marco Avesani$^1$, \\
Tommaso Bertapelle$^1$, 
Daniele Dequal$^3$, 
Giuseppe Vallone$^{1,4}$ 
and Paolo Villoresi$^1$ \\
 \small{ \textit{$^1$Dipartimento di Ingegneria dell'Informazione, Universit\`a degli Studi di Padova, via Gradenigo 6B, 35131 Padova, Italy}}\\
 \small{  \textit{$^2$Dipartimento di Matematica "Tullio Leci-Civita", Universit\`a degli Studi di Padova, Via Trieste 63, 35121 Padova, Italy}}\\
 \small{ \textit{$^3$Unit\`a Telecomunicazioni e Navigazione, Agenzia Spaziale Italiana, contrada Terlecchia s.n.c., Matera, Italy}}\\
 \small{ \textit{$^4$Dipartimento di Fisica e Astronomia, Universit\`a degli Studi di Padova, via Marzolo 8, 35131 Padova, Italy }}
}

\maketitle
\copyrightnotice
\footnotetext[1]{andrea.stanco@unipd.it}
\begin{abstract}
This work presents a hardware and software architecture which can be used in those systems that implement practical Quantum Key Distribution (QKD) and Quantum Random Number Generation (QRNG) schemes. This architecture fully exploits the capability of a System-on-a-Chip (SoC) which comprehends both a Field Programmable Gate Array (FPGA) and a dual core CPU unit. By assigning the time-related tasks to the FPGA and the management to the CPU, we built a flexible system with optimized resource sharing on a commercial off-the-shelf (COTS) evaluation board which includes a SoC. Furthermore, by changing the dataflow direction, the versatile system architecture can be exploited as a QKD transmitter, QKD receiver and QRNG control-acquiring unit. Finally, we exploited the dual core functionality and realized a concurrent stream device to implement a practical QKD transmitter where one core continuously receives fresh data at a sustained rate from an external QRNG source while the other operates with the FPGA to drive the qubits transmission to the QKD receiver. The system was successfully tested on a long-term run proving its stability and security. This demonstration paves the way towards a more secure QKD implementation, with fully unconditional security as the QKD states are entirely generated by a true random process and not by deterministic expansion algorithms. Eventually, this enables the realization of a standalone quantum transmitter, including both the random numbers and the qubits generation.
\end{abstract}


\section{Introduction}
\label{sec:introduction}
\IEEEPARstart{Q}{uantum} Communication (QC) is one of the promising applications of quantum technology and is recently receiving a relevant boost towards commercial applications. Quantum Key Distribution (QKD) and Quantum Random Number Generation (QRNG) are the two leading technologies of QC since their combination allows to realize the perfect secrecy protocol, resistant to any external attack. The realization of such a system requires the design and development of several components: from the optical setup to the driving electronics, from the digital control board to the management software. Besides the particular quantum implementations which can vary according to different security protocols, an essential component of the whole setup is a supervision board capable to provide deterministic behaviour, high temporal resolution, and high speed computation. A Field Programmable Gate Array (FPGA) is almost a mandatory choice for such applications \cite{Bacco2013, Wei2020, Zhang2012}. FPGA also offers an advantage in terms of power consumption \cite{Qasaimeh2019}, which can be a key feature for critical applications such as Cubesat missions for Satellite Quantum Communication \cite{Oi2017}. In this work, we present a generalized FPGA-based architecture that can be easily applied to Discrete Variable QKD (DV-QKD) and QRNG (DV-QRNG) applications. Moreover, the schematic has the potentiality to be used even in Continuous Variables QKD (CV-QKD) and QRNG (CV-QRNG), provided that an auxiliary Digital-to-Analog (DAC) and Analog-to-Digital (ADC) should be included. The architecture was implemented and tested on the Zynq-7020 System-on-a-Chip (SoC) mounted on an entry level evaluation board: ZedBoard by Avnet. The exploitation of both the FPGA layer and the CPU one of the SoC leads to a high level of flexibility, allowing to scale the application functionalities to the specific part of the chip. According to the specific application, the system can be set in a top-down (dataflow from pc/user to quantum system) or bottom-up (dataflow from quantum system to pc/user) configuration. The system was successfully used in different configurations in several experiments over the past years \cite{Hamid2021, Stanco2020, Avesani2021_2, iPognac, Calderaro2020, Agnesi2020, Avesani2021, Pognac2019, Vedovato2017}. Recently, it was also tested in the prototype of a QKD transmitter for Cubesat mission \cite{Balossino2020} making it suitable for Satellite Quantum Communications, a key sector of QC. Given the presence of a dual core CPU, we also designed, developed, and successfully tested a dual core application capable to sustain a continuous data transfer from an external source to the CPUs and then to the FPGA. This feature is the key to implement a provably secure QKD system since it allows to combine a QRNG output stream with a QKD stream without the need of random number expansion \cite{Walenta2014, Constantin2017}, paving the way to commercial QKD devices with full unconditional security.
The work is structured as follow: in Section \ref{sec:archOver} an overall overview of the architecture is presented; in Sections \ref{sec:fpgaLayer} and \ref{sec:cpuLayer} the FPGA and CPU layers are described; in Section \ref{sec:stream} the dual core architecture for a QKD transmitter is presented along with the system test results.

\section{Architecture Overview}
\label{sec:archOver}
\begin{figure}[t!]
\centering
\includegraphics[scale=0.51]{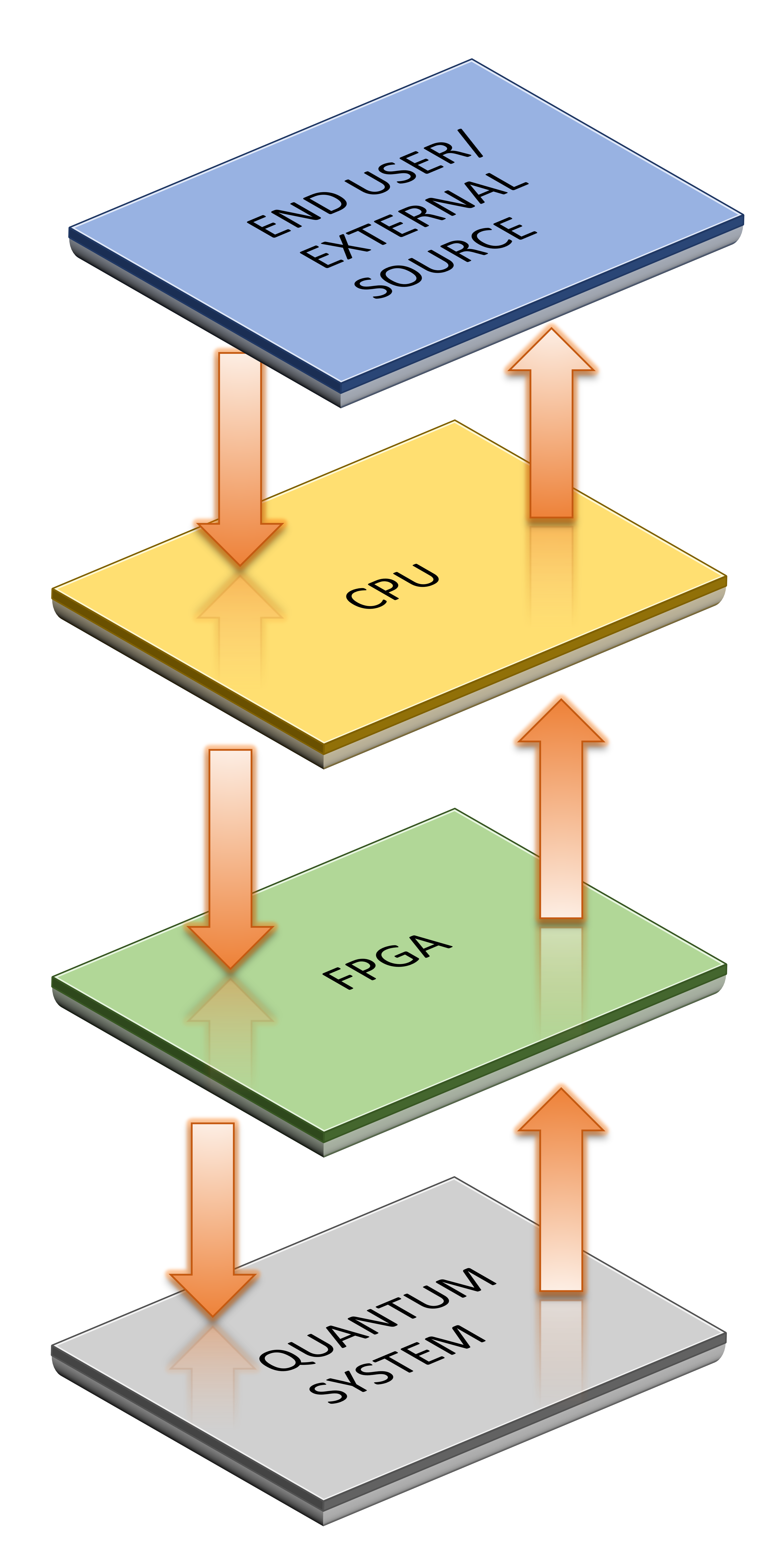}
\caption{Overview of the four-layer structure of the system. The embedded architecture is divided into two different layers (FPGA+CPU) which are enclosed by the two outside-world layers (End User/External Source and Quantum System).\label{Main_Scheme}}
\end{figure}

The system architecture is organized in two different layers. The lower layer is the FPGA one where all the deterministic and high resolution operations are carried out. The higher layer is the CPU(s) one which is responsible for the parameters and data management operations as well as the communication with the outside world. Besides the functions separation, this subdivision is also a key point for the maintenance and the upgrading of the architecture since the two layers require different programming languages (VHDL/Verilog vs. C/C++) and different design teams. Two additional layers, End User/External Source and Quantum System, enclose the previous ones completing the whole practical system. The architecture is designed to easily switch between top-down and bottom-up workflows, which represents the distinction between QKD transmitter and QKD receiver/QRNG applications.
An overview of the whole schematic is given in Figure \ref{Main_Scheme}. In the following, we provide a general description of the Top-Down and Bottom-Up applications.

\subsection{Top-Down application}
In this configuration, the dataflow starts from an external device, e.g., a QRNG or PC, goes through the CPU, then to the FPGA and finally to the chip input-output pins (I/O). This layout is suitable for a QKD transmitter, as the raw cryptographic key, either generated in real time by a QRNG or previously stored in a PC, is fed through the CPU to the FPGA, which drives the hardware dedicated to quantum state generator accordingly. As detailed in the following, the first communication step, from the external device to the CPU, has been performed over Gigabit Ethernet. This choice provides both a high throughput of the data transfer (>600 Mbit/s) and a great flexibility, being Ethernet a widespread standard. As there is no encryption of the dataflow, it is of paramount importance to protect the communication channel from eavesdropping. This can be done by setting up a private Local Area Network (LAN), physically disconnected from other networks, between the SoC and the external device. The second step in the stack is from the CPU to the FPGA. For this step, two solutions have been implemented. The configuration parameters, e.g., qubit frequency or total transmission length, which have a very slow refresh rate, are exchanged with a direct communication via the Advanced eXtensible Interface (AXI) protocol. Instead, for the raw key exchange, which can reach 400 Mbps of steady data transmission, we exploited both the onboard DDR-RAM memory, accessible from the CPU, and the Block RAM memory (BRAM), integrated in the FPGA but accessible from the CPU through AXI protocol. The BRAM, with a maximum length in the order of Mbits, has been divided in two halves, so that while the FPGA is reading from one section, the CPU can update the contents of the other with the data stored in the RAM and previously received via Ethernet. This allows for a continuous and synchronized dataflow from the external device to the FPGA, and hence its I/Os. According to the specific quantum system, the output signals are routed to either PMOD (LVCMOS33 standard) or FMC ports (LVCMOS18 standard) of the ZedBoard and then properly amplified by an external driving stage.

\subsection{Bottom-Up application}
In this configuration, the dataflow starts from the chip I/Os controlled by the FPGA, is then transmitted from the FPGA to the CPU and finally from the CPU to the external device. 
This layout can be used either for a QKD receiver or for a QRNG, where the electrical signal coming from external devices, i.e., single photon detectors, is sampled by the FPGA I/Os. The sampled and stored signal is then transferred, via the CPU, to an external computer, for the implementation of the post-processing phases of the QKD protocol, i.e., parameter estimation, error correction and privacy amplification. The communication interfaces for the bottom-up configuration are the same as for the top-down. 
It must be mentioned that also in this case the communication between the CPU and the external computer must be performed over a secure LAN, as the data stream at this level is not encrypted.
To guarantee a high level of security, the LAN used for the PC-CPU communication has to be reserved and so physically separated from the one used for communication between the transmitter and receiver PCs.

\section{FPGA Layer}
\label{sec:fpgaLayer}
\begin{figure}[t]
\includegraphics[scale=0.43]{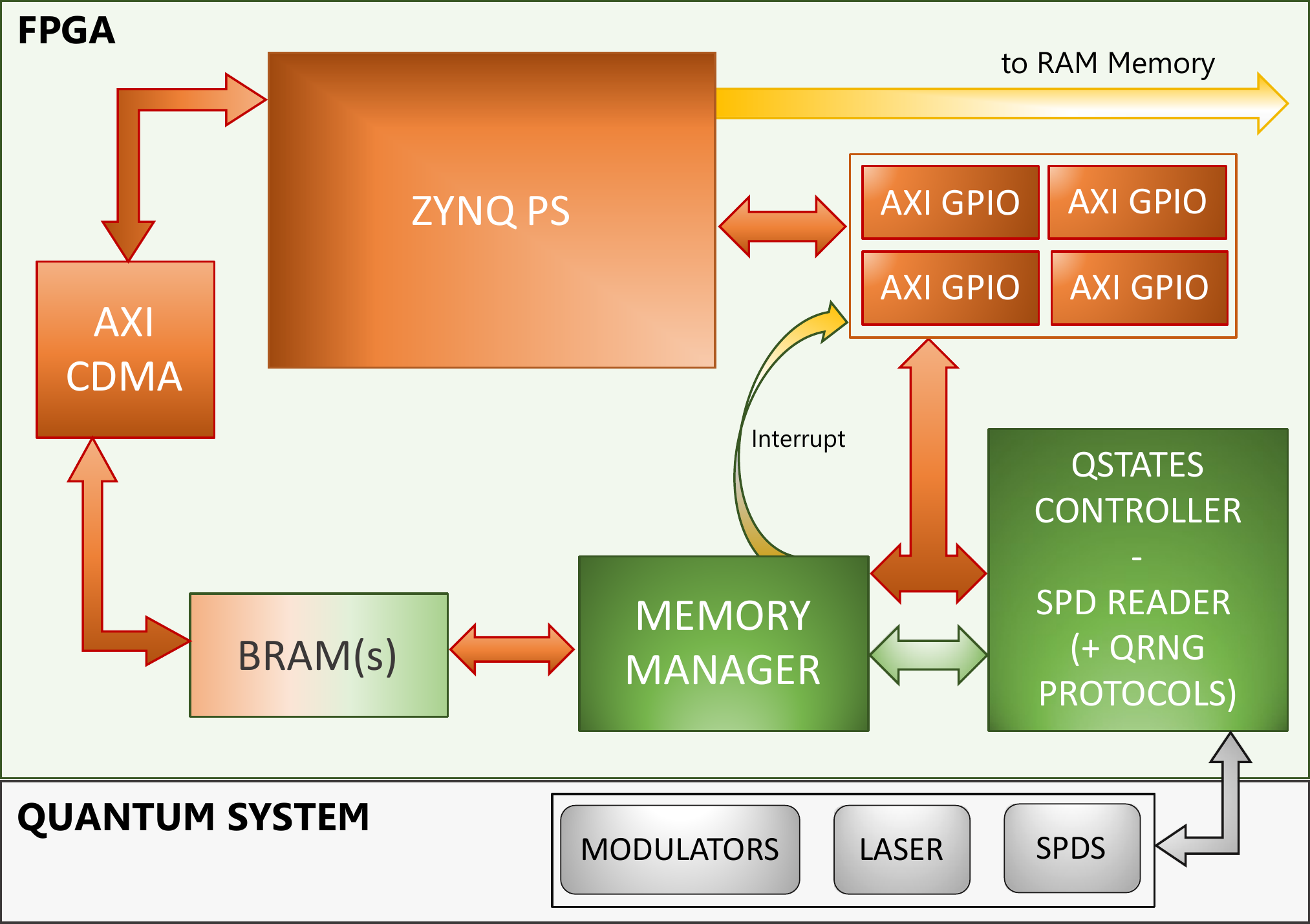}
\caption{Schematic view of the FPGA system. The modules in green are custom VHDL blocks which control the memory management, the qubit generation and the single photon detector readout. The orange modules are the AXI-based ones and they are AXI-CDMA and AXI-GPIO. The BRAM module has a mixed color since it is controlled both by AXI and by custom VHDL modules. The CPU part is initialized in the FPGA design and is identified by the ZYNQ Processing System (PS) module which can also access to RAM memory. \label{FPGA_Scheme}}
\end{figure}

The FPGA implementation allows to have a perfect time control over the optical system. Indeed, the capability to schedule every operation according to a system clock is a key feature for the realization of a QKD/QRNG system. When used in a QKD transmitter configuration, the FPGA is responsible for the rightful generation of the electrical pulses which drive the electro-optical elements of the setup. When used as a QRNG/QKD receiver, the FPGA takes care of the read-out operations of the external electrical signals coming from single-photon detectors. For the sake of completeness, it is also possible to consider the application of the architecture for CV-QKD \cite{Laudenbach2018}  and CV-QRNG \cite{Marangon2017,Avesani2018}. The difference would be that the FPGA needs to interface with proper external DAC (for CV-QKD transmitter) and external ADC (for CV-QKD receiver and CV-QRNG).
The general and simplified structure of the FPGA design is shown in Figure \ref{FPGA_Scheme}. The design uses AXI-capable blocks for communication and data transfer to (from) the CPU along with BRAMs and custom VHDL blocks for FPGA data management, Memory Manager (MM) block, and external signal generation (readout), QStates Controller (QSC) and SPD Reader (SR) blocks. The QRNG application also includes dedicated modules implementing random generation protocols \cite{Furst2010,Stipc2007}. AXI-GPIOs allow to set parameters from the CPU and to read out interrupt signals asserted from custom VHDL blocks. AXI-CDMA enables the possibility to move data to (from) the board RAM from (to) the BRAM(s).
The MM is responsible for managing  the data transfer between the BRAM(s) and the other custom blocks. Being the BRAM divided into two halves, the MM asserts a signal every time it reaches the end of one of the two halves (while reading or writing) and, in turn, the signal is read by an AXI-GPIO and interpreted as an interrupt from the CPU which writes (reads) new data to (from) the BRAM.
The system clock was set in a range between 100 and 200 MHz depending on the specific application. Future improvements will consider pushing further this frequency in order to increase the overall system speed. Indeed, apart from tighter timing constraints, a higher frequency clock implies a higher data throughtput to/from the external device, which may exceed the gigabit range of the board.

\subsection{QKD Transmitter}
\label{sec:qkdtx}
In addition to the previously described general architecture, the FPGA for a DV-QKD transmitter requires a specific design of the QSC module to encode the raw key data into electrical output pulses which drive the laser and electro-optical modulators. We designed several variations of this module according to the chosen QKD protocol \cite{Tamaki2014, Hwang2003, Lo2005,Rusca2018}, derived from the well-known BB84 \cite{Bennett2014_BB84}, and implementation \cite{iPognac, Pognac2019, Agnesi2020, Balossino2020}. Here, we give a brief description of one of the most recent versions, presented in \cite{Balossino2020}. With a system clock set to 200 MHz, a pulse of 5 ns can be provided at the output. The encoding of every qubit requires, nominally, no more than 15 ns time slot since the polarization encoding describes three different polarization states and necessitates an output pulse in three different temporal positions (0-5 ns, 5-10 ns, 10-15 ns). The decoy implementation works in a similar way, describing three different intensity level, but requests an output pulse in only two temporal position (0-5 ns, 5-10 ns) in combination with a possible laser switch off. The pulse for the laser driver is output at the begin of every slot (0-5 ns) unless the case of a specific decoy state. To compensate and synchronize the output signals with the optical path length, a dedicated time offset can also be applied to each signal.
Two different BRAMs were instantiated, one for the polarization encoding data and one for the decoy one. Since every qubit requires two bits to distinguish among three polarization states as well as other two bits to discriminate among three decoy states, the BRAMs were set to the same size and operates with the same interrupt routine. For the sake of clearness, it is possible to optimize the overall qubit encoding using just three bits for the polarization+decoy encoding. Nevertheless, we chose to use two+two encoding for mainly two reasons. The first reason is that a three-bit encoding would have required a quite complicated, and in a certain way inefficient, routine to distinguish the data within each byte as more than two but less than three-qubits-encoding-data would have been stored in one byte. The second reason is that a two+two bit encoding allows to separate the paths, the memories and, in turn, the TCP sockets of the polarization and decoy data enhancing the robustness and flexibility of the overall system.

\subsection{QKD Receiver/QRNG}
\label{sec:qkdrx}
In principle, the FPGA schematic for a QKD receiver is similar to the one for a DV-QRNG. In both cases, the I/Os are connected to the output signals of single-photon detectors and the FPGA implements the sampling process to produce a bitstring containing the digital temporal description of the single-photon events. First of all, the input signal is translated to the FPGA clock domain by using a proper async-to-sync hardware module included in the SR. Then, in the case of a QRNG application, the sampled bits are temporarily accumulated and later processed by proper modules which apply random generation protocols to a small set of data, as described in \cite{Stanco2020}. The random bit is then stored and managed by the MM which, in turn, transfers a 32 bit array to the i-th address of a BRAM and calls an interrupt whenever it reaches half or the end of it. Moreover, this architecture was also the perfect option to implement a synchronized QRNG which was needed for the realization of \cite{Vedovato2017}. For this application, which required to output a random number generated only after a specific trigger event, the architecture was modified to allow a resetting of all the random data (even the sampled bits). The resetting was triggered by an external electrical signal coming from the experiment setup. The random bit was then used to produce an auxiliary output port which set a component of the experiment. In order to improve the randomness of the output number, the architecture was also doubled and produced two random bits which were XORed. For further details, refer to \cite{Vedovato2017}. As a matter of fact, by removing the generation protocol modules, the architecture becomes suitable to be used as a QKD receiver. However, a drawback of this implementation is the low time resolution provided by the system clock which, even in a high-range FPGA-chip case scenario, does not exceed 1 GHz. For a high performance QKD a sub-ns time resolution at the receiver is required. Therefore, this implementation can be a solution only for low cost QKD systems. Nevertheless, the design and integration of a Time-to-Digital Converter (TDC) FPGA-module (such as \cite{Song2006, Fishburn2013, Chen2019}) or the exploitation of an external integrated circuit (such as \cite{TDC-GPX}) would allow for a sub-ns time resolution and thus the use in high performance QKD systems. Indeed, future steps will investigate such solutions. 

\section{CPU Layer}
\label{sec:cpuLayer}
The CPU software is implemented as a standalone/bare-metal application and no Operative System is required. This has the great advantage of having a very light and fast software at a higher design cost. The software, developed in C and C++ languages, has the role to interface the FPGA layer with the external source and the final user. It mainly implements the interrupt routine to move (read) data from (to) the RAM to (from) the BRAM anytime the MM reaches the half of the end of a BRAM. It also implements the TCP connection sockets to receive commands and data from the external source or user.

\subsection{TCP Connection}
To communicate with the outside world, a TCP protocol was chosen. Given the robustness of the TCP protocol over any possibility of losing data packets, this choice has to be preferred over the UDP protocol which cannot guarantee a reliable communication to the application layer, thus undermining the validity of the QKD implementation. For a QRNG application where the data are output to an external receiver, a UDP protocol might be suitable in any case since any (negligible) data losses do not affect the overall quality and performance of the QRNG device. Nevertheless, for QKD post-processing purposes, one can consider to send the random stream to two different devices, e.g., a QKD transmitter and a computer, and in this case a data loss would jeopardize the whole system. Moreover, changing the protocol only for the QRNG application would reduce the overall system flexibility.

The connection between the PC and the CPU applications is structured in two or more different TCP server-client sockets: one is for commands and parameters exchange while the other(s) is for data exchange. Indeed, the data socket has the only purpose to receive new data from the external source. Thus, the number of required operations and statement conditions in the \textit{data-received-callback} are quite few allowing to have optimal performance over the TCP bandwidth (>600 Mbit/s).

\section{QKD-Tx Continuous Stream Implementation}
\label{sec:stream}

\begin{figure*}[t!]
\includegraphics[scale=0.87]{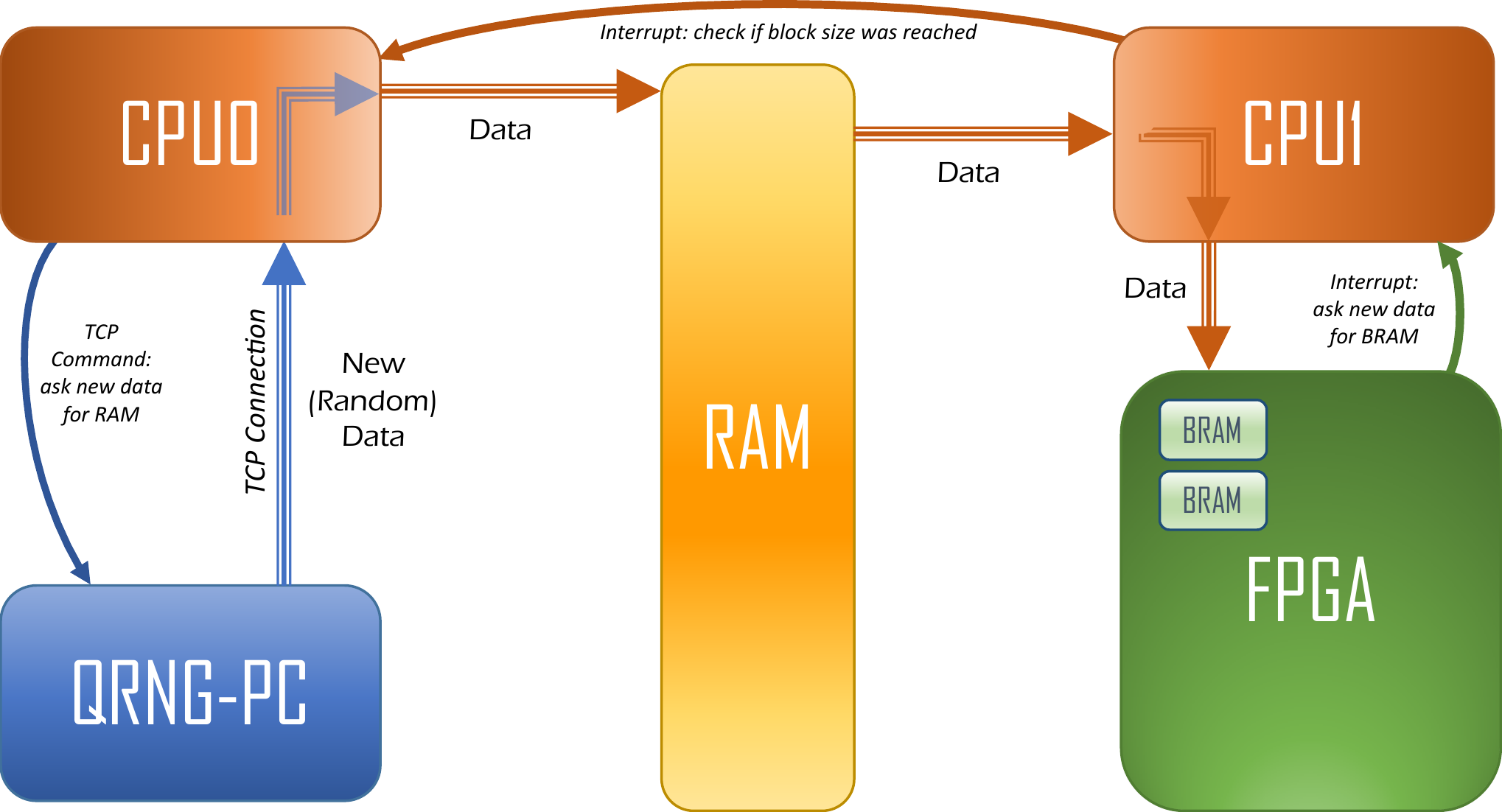}
\caption{Schematic view of the dual core system, representing the flow of data from the QRNG-PC to the FPGA. The request of new data is triggered by the FPGA, each time it reads half of the BRAM, by means of an interrupt to the CPU1, which then moves the data from the buffer to the RAM. Each time CPU1 reads 18.75~MiB of data from the buffer it sends an interrupt to the CPU0 to update the new block with fresh data from the QRNG-PC.\label{Dual-Core_Scheme}}
\end{figure*}

A fundamental feature of any future commercial QC device is the capability to continuously sustain the transmission of fresh random data. That is, the external randomness source, like a QRNG device or a computer where random keys are stored, must provide new data with a sufficient bitrate and the QKD transmitter must be able to perform at the same time both the reading/storing of these data and the transmission of electrical pulses. Nowadays, specific workarounds allow to avoid the implementation of such functionality \cite{Walenta2014} but necessarily reduce the overall security of the system. For instance, one can connect a low bitrate QRNG to the QKD source where the random data is expanded to reach the required bitrate \cite{Constantin2017}. The expansion process, although implemented via standard cryptographic primitives, does not offer an unconditional type of security, and can represent a security breach in the entire QKD system.
Hence, we developed a dual core architecture able to sustain the required data rate for a secure QKD implementation, allowing a random data stream generated entirely by a QRNG. This approach has the advantage of being unconditionally secure. Moreover, keeping the random data stored on the PC eases the QKD postprocessing procedure or, alternatively, lowers the memory resources of the SoC needed to store the transmitted bitstring until the receiver communicates the detected qubits. Furthermore, the bias of the random bits, required by some efficient QKD protocols \cite{Lo2005}, can be adjusted without the need of programming or setting the FPGA, allowing to optimize it according to the current quantum channel condition.

The data stream flow is represented in Figure~\ref{Dual-Core_Scheme}. The data generated from the QRNG-PC is received by a TCP socket and then moved by the CPU0 into a buffer in the RAM. Meanwhile, the CPU1 reads the data from the RAM and moves it to the BRAM, which can be read by the FPGA. The buffer size is set to 187.5~MiB and divided into 10 blocks, that are written atomically by the CPU0. Hence, when a block has been moved to the FPGA, an interrupt from the CPU1 is sent to the CPU0 to notify that a new block of the buffer can be written. The CPU0 forward a request of a new block of 18.75~MiB to the QRNG-PC. The CPU1 reads smaller chuncks from the buffer, whose size is half of the BRAM's size, when it receives an interrupt from the FPGA. Compared to the BRAM size, the buffer is larger to avoid an unwanted stop of the continuous feed of data to the FPGA, which may happen due to the temporary loss of speed of the TCP channel, or the latency of the CPU0. The whole architecture is doubled in order to manage both the stream for polarization and decoy data.

\subsection{System tests}
We implemented this system on a QKD transmitter to test the top-down application in continuous stream mode. We performed a double stress test: the first one was with a real QRNG device, based on the scheme of \cite{Gabriel2010a} offering high security and bitrate, while the second test was carried out with a Cryptographically Secure Pseudo Random Number Generator (CSPRNG) able to provide a sufficient data rate as well. We chose to perform the test also with a CSPRNG to show the system capability in a fallback scenario where no QRNG device is available. The (pseudo-)random data was stored in a buffer of the PC, to ease the retrieve of the quantum states sent to the receiver needed for the raw key sifting. Indeed, once the QKD transmitter has produced and sent the quantum states, the QKD receiver measures and detects a subset of these states due to unavoidable channel losses. The QKD receiver communicates to the transmitter the list of states it detected (without revealing the outcome of the measurement). Hence, the QKD transmitter selects the subset of random data that will be used for the sifting.

\begin{figure}[h!]
\centering
\includegraphics[scale=0.57]{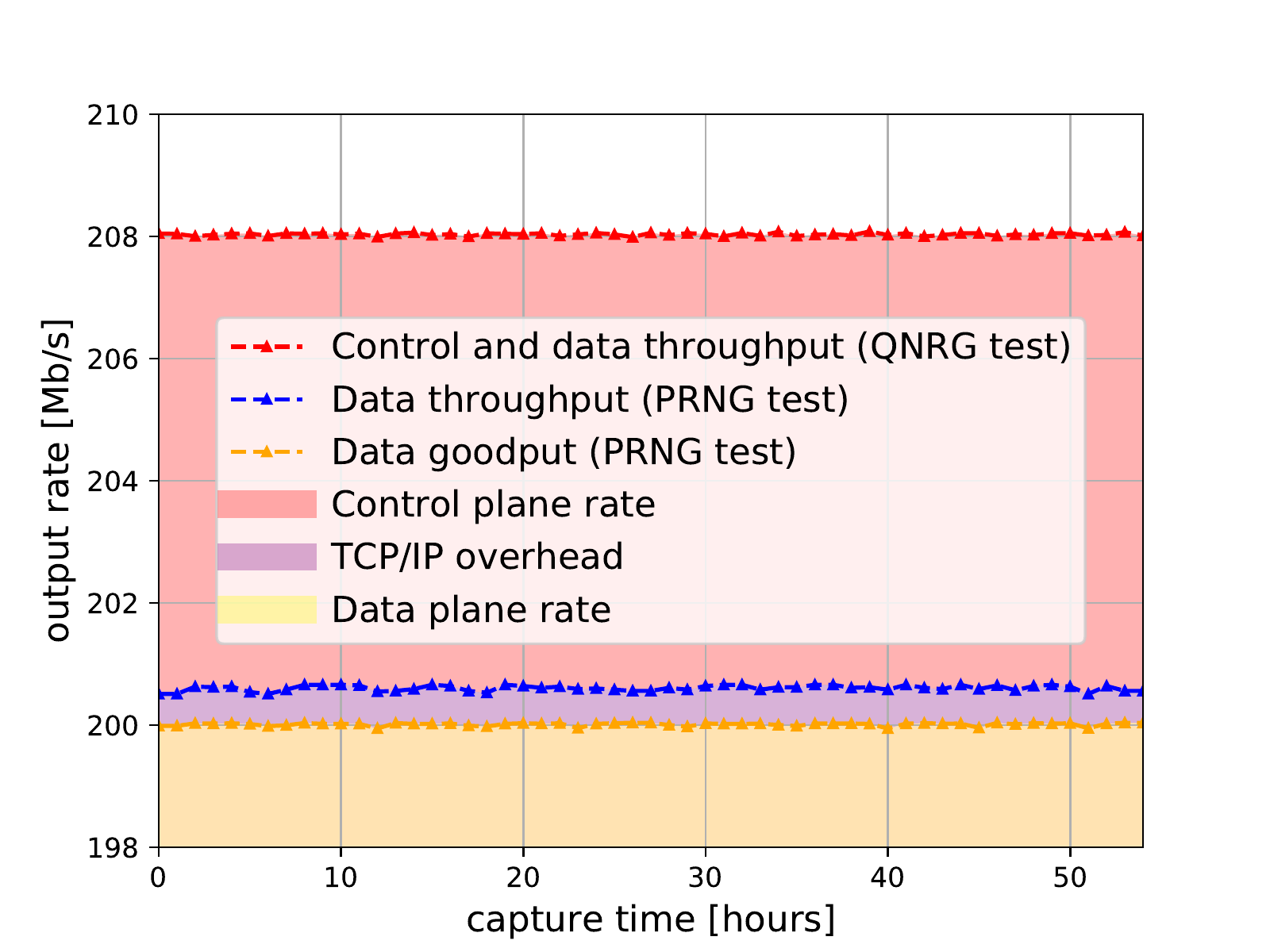}
\caption{Plot of the TCP traffic from the PC to the Zynq-7020 for the transmission of the random sequences and control messages, for two 55 hours long tests using either the QRNG or CSPRNG as randomness source. We plotted in red the aggregated traffic of the data and control plane for the QRNG test, amounting totally to about 208 Mb/s. The transmission of states and intensities sequences accounts for 200 Mb/s. This stream is represented in yellow with the data from the PRNG test. Finally, we can appreciate a small TCP/IP overhead (as seen from the PC OS) due to TCP segmentation offload. \label{throughput_plot}}
\end{figure}

The stream of data at the PC is managed by three threads: one for the production of blocks of random sequences;  one for the system managing the transfer, upon request, of the random data from the buffer to the board \textit{via} TCP; and one for the selection of the states detected by the QKD receiver.
As anticipated, the (pseudo-)random data can either be received from a QRNG device or generated internally by a CSPRNG. In the former case, the first thread is used to receive (\textit{via} UDP or TCP) the random bits from the QRNG and to bias them according to the desired Bernoulli distribution. In the latter case, the thread carry on the generation of the pseudo-random data by using a Chacha20 based CSPRNG, seeded with the Intel® Secure Key hardware random number generator embedded in the recent generations of Intel CPUs.
We synchronized the write and read operations on the buffer of these multiple threads by  using semaphores. The buffer was divided in chuncks that could be either written or read at a time. Hence, one semaphore is needed to allow the writing of new chuncks of data by the RNG thread, which can be done only on those chuncks that have been read by the thread selecting the subset of states arrived at the QKD receiver. Another semaphore is needed to ensure that the chuncks of data sent to the system are those that have been rewritten by the RNG thread. 

In our test, the QKD repetition rate was set to 50~MHz. Since the state encoding uses two bits for the state polarization and two bits for the mean photon number, we have two data streams of 100~Mb/s, plus a third stream for control communications. The outbound traffic was monitored from the transmitting PC during the two tests and is reported in Figure~\ref{throughput_plot}. Given this steady input, the system was able to carry on all the needed operations seamlessly along all the 55 hours of the tests, resulting in a successful execution of the QKD protocol.

\section{Conclusion}
In this work, we presented a versatile architecture based on FPGA technology exploiting also a CPU counterpart for the implementation of practical Quantum Communication systems. The architecture was developed in different layers with different tasks and is easily interchangeable among different QC applications such DV-QKD transmitter, DV-QKD receiver and DV-QRNG. We also implemented and tested a dual core functionality performing a TCP continuous stream between a QRNG source and a QKD transmitter without the need of data expansion to reach the amount of data required to encode every qubit. This allows to strengthen the security of the QKD implementation, as the random settings needed by the QKD protocol are guaranteed to be unconditionally secure, unlike those generated by expansion algorithms. The system was implemented on a low-budget COTS and it was successfully tested to continuously provide 4 bits to encode a qubit every 20 ns. Future steps will consider higher frequency implementation as well as CV applications by including proper DAC and ADC hardwares.

\section*{Acknowledgment}

Part of this work was supported by Ministero dell’Istruzione, dell’Università e della Ricerca (MIUR) (Italian Ministry of Education, University and Research) under the initiative “Departments of Excellence” (Law 232/2016). A. S. and D. D. Authors would like to thank Dr. S. Gaiarin for useful help and hints in the general architecture design. A. S. and L. C. Authors would like to thank Mr. F. Berra for useful discussions about dual core architecture verification. A. S. Author would like to thank Dr. D. G. Marangon for useful help and discussion for the DV-QRNG architecture design. 

\bibliographystyle{unsrt}
\bibliography{bib_resource}

\begin{thebibliography}{10}

\bibitem{Bacco2013}
Davide Bacco, Matteo Canale, Nicola Laurenti, Giuseppe Vallone, and Paolo
  Villoresi.
\newblock Experimental quantum key distribution with finite-key security
  analysis for noisy channels.
\newblock {\em Nature Communications}, 4(1):2363, Sep 2013.

\bibitem{Wei2020}
Kejin Wei, Wei Li, Hao Tan, Yang Li, Hao Min, Wei-Jun Zhang, Hao Li, Lixing
  You, Zhen Wang, Xiao Jiang, Teng-Yun Chen, Sheng-Kai Liao, Cheng-Zhi Peng,
  Feihu Xu, and Jian-Wei Pan.
\newblock High-speed measurement-device-independent quantum key distribution
  with integrated silicon photonics.
\newblock {\em Phys. Rev. X}, 10:031030, Aug 2020.

\bibitem{Zhang2012}
H.~F. {Zhang}, J.~{Wang}, K.~{Cui}, C.~L. {Luo}, S.~Z. {Lin}, L.~{Zhou},
  H.~{Liang}, T.~Y. {Chen}, K.~{Chen}, and J.~W. {Pan}.
\newblock A real-time qkd system based on fpga.
\newblock {\em Journal of Lightwave Technology}, 30(20):3226--3234, 2012.

\bibitem{Qasaimeh2019}
Murad Qasaimeh, Kristof Denolf, Jack Lo, Kees Vissers, Joseph Zambreno, and
  Phillip~H. Jones.
\newblock Comparing energy efficiency of cpu, gpu and fpga implementations for
  vision kernels.
\newblock In {\em 2019 IEEE International Conference on Embedded Software and
  Systems (ICESS)}, pages 1--8, 2019.

\bibitem{Oi2017}
Daniel~KL Oi, Alex Ling, Giuseppe Vallone, Paolo Villoresi, Steve Greenland,
  Emma Kerr, Malcolm Macdonald, Harald Weinfurter, Hans Kuiper, Edoardo
  Charbon, and Rupert Ursin.
\newblock Cubesat quantum communications mission.
\newblock {\em EPJ Quantum Technology}, 4(1):6, Apr 2017.

\bibitem{Hamid2021}
Hamid Tebyanian, Mujtaba Zahidy, Marco Avesani, Andrea Stanco, Paolo Villoresi,
  and Giuseppe Vallone.
\newblock Practical semi-device independent randomness generation based on
  quantum state's indistinguishably, 2021.

\bibitem{Stanco2020}
Andrea Stanco, Davide~G. Marangon, Giuseppe Vallone, Samuel Burri, Edoardo
  Charbon, and Paolo Villoresi.
\newblock Efficient random number generation techniques for cmos single-photon
  avalanche diode array exploiting fast time tagging units.
\newblock {\em Phys. Rev. Research}, 2:023287, Jun 2020.

\bibitem{Avesani2021_2}
Marco Avesani, Luca Calderaro, Giulio Foletto, Costantino Agnesi, Francesco
  Picciariello, Francesco B.~L. Santagiustina, Alessia Scriminich, Andrea
  Stanco, Francesco Vedovato, Mujtaba Zahidy, Giuseppe Vallone, and Paolo
  Villoresi.
\newblock Resource-effective quantum key distribution: a field trial in padua
  city center.
\newblock {\em Opt. Lett.}, 46(12):2848--2851, Jun 2021.

\bibitem{iPognac}
Marco Avesani, Costantino Agnesi, Andrea Stanco, Giuseppe Vallone, and Paolo
  Villoresi.
\newblock Stable, low-error, and calibration-free polarization encoder for
  free-space quantum communication.
\newblock {\em Opt. Lett.}, 45(17):4706--4709, Sep 2020.

\bibitem{Calderaro2020}
Luca Calderaro, Andrea Stanco, Costantino Agnesi, Marco Avesani, Daniele
  Dequal, Paolo Villoresi, and Giuseppe Vallone.
\newblock Fast and simple qubit-based synchronization for quantum key
  distribution.
\newblock {\em Phys. Rev. Applied}, 13:054041, May 2020.

\bibitem{Agnesi2020}
Costantino Agnesi, Marco Avesani, Luca Calderaro, Andrea Stanco, Giulio
  Foletto, Mujtaba Zahidy, Alessia Scriminich, Francesco Vedovato, Giuseppe
  Vallone, and Paolo Villoresi.
\newblock Simple quantum key distribution with qubit-based synchronization and
  a self-compensating polarization encoder.
\newblock {\em Optica}, 7(4):284--290, Apr 2020.

\bibitem{Avesani2021}
M.~Avesani, L.~Calderaro, M.~Schiavon, A.~Stanco, C.~Agnesi, A.~Santamato,
  M.~Zahidy, A.~Scriminich, G.~Foletto, G.~Contestabile, M.~Chiesa, D.~Rotta,
  M.~Artiglia, A.~Montanaro, M.~Romagnoli, V.~Sorianello, F.~Vedovato,
  G.~Vallone, and P.~Villoresi.
\newblock Full daylight quantum-key-distribution at 1550 nm enabled by
  integrated silicon photonics.
\newblock {\em npj Quantum Information}, 7(1):93, Jun 2021.

\bibitem{Pognac2019}
Costantino Agnesi, Marco Avesani, Andrea Stanco, Paolo Villoresi, and Giuseppe
  Vallone.
\newblock All-fiber self-compensating polarization encoder for quantum key
  distribution.
\newblock {\em Opt. Lett.}, 44(10):2398--2401, May 2019.

\bibitem{Vedovato2017}
Francesco Vedovato, Costantino Agnesi, Matteo Schiavon, Daniele Dequal, Luca
  Calderaro, Marco Tomasin, Davide~G. Marangon, Andrea Stanco, Vincenza Luceri,
  Giuseppe Bianco, Giuseppe Vallone, and Paolo Villoresi.
\newblock Extending wheeler{\textquoteright}s delayed-choice experiment to
  space.
\newblock {\em Sci. Adv.}, 3(10):e1701180, 2017.

\bibitem{Balossino2020}
A.~Balossino, E.~Fazzoletto, S.~Ciaglia, N.~Tisat, A.~Di~Paola, E.~Scarpa,
  B.~Cotugno, C.~Agnesi, L.~Calderaro, A.~Stanco, G.~Vallone, and P.~Villoresi.
\newblock Seqbo - a miniaturized system for quantum key distribution.
\newblock volume 2020-October, 2020.

\bibitem{Walenta2014}
N.~Walenta, A.~Burg, D.~Caselunghe, J.~Constantin, N.~Gisin, O.~Guinnard,
  R.~Houlmann, P.~Junod, B.~Korzh, N.~Kulesza, M.~Legr{\'{e}}, C.~W. Lim,
  T.~Lunghi, L.~Monat, C.~Portmann, M.~Soucarros, R.~T. Thew, P.~Trinkler,
  G.~Trolliet, F.~Vannel, and H.~Zbinden.
\newblock A fast and versatile quantum key distribution system with hardware
  key distillation and wavelength multiplexing.
\newblock {\em New Journal of Physics}, 16(1):013047, jan 2014.

\bibitem{Constantin2017}
Jeremy Constantin, Raphael Houlmann, Nicholas Preyss, Nino Walenta, Hugo
  Zbinden, Pascal Junod, and Andreas Burg.
\newblock An fpga-based 4 mbps secret key distillation engine for quantum key
  distribution systems.
\newblock {\em Journal of Signal Processing Systems}, 86(1):1--15, Jan 2017.

\bibitem{Laudenbach2018}
Fabian Laudenbach, Christoph Pacher, Chi-Hang~Fred Fung, Andreas Poppe,
  Momtchil Peev, Bernhard Schrenk, Michael Hentschel, Philip Walther, and
  Hannes H{\"{u}}bel.
\newblock {Continuous-Variable Quantum Key Distribution with Gaussian
  Modulation-The Theory of Practical Implementations}.
\newblock {\em Advanced Quantum Technologies}, 1(1):1800011, aug 2018.

\bibitem{Marangon2017}
Davide~G. Marangon, Giuseppe Vallone, and Paolo Villoresi.
\newblock Source-device-independent ultrafast quantum random number generation.
\newblock {\em Phys. Rev. Lett.}, 118:060503, 02 2017.

\bibitem{Avesani2018}
Marco Avesani, Davide~G. Marangon, Giuseppe Vallone, and Paolo Villoresi.
\newblock {Source-device-independent heterodyne-based quantum random number
  generator at 17 Gbps}.
\newblock {\em Nature Communications}, 9(1):5365, dec 2018.

\bibitem{Furst2010}
Harald F\"{u}rst, Henning Weier, Sebastian Nauerth, Davide~G. Marangon,
  Christian Kurtsiefer, and Harald Weinfurter.
\newblock High speed optical quantum random number generation.
\newblock {\em Opt. Express}, 18(12):13029--13037, 06 2010.

\bibitem{Stipc2007}
M.~Stip\v{c}evi\'{c} and B.~Medved Rogina.
\newblock Quantum random number generator based on photonic emission in
  semiconductors.
\newblock {\em Rev. Sci. Instrum.}, 78(4):045104, 2007.

\bibitem{Tamaki2014}
Kiyoshi Tamaki, Marcos Curty, Go~Kato, Hoi-Kwong Lo, and Koji Azuma.
\newblock {Loss-tolerant quantum cryptography with imperfect sources}.
\newblock {\em Physical Review A}, 90(5):052314, nov 2014.

\bibitem{Hwang2003}
Won-Young Hwang.
\newblock Quantum key distribution with high loss: Toward global secure
  communication.
\newblock {\em Phys. Rev. Lett.}, 91:057901, Aug 2003.

\bibitem{Lo2005}
Hoi-Kwong Lo, H.~F. Chau, and M.~Ardehali.
\newblock Efficient quantum key distribution scheme and a proof of its
  unconditional security.
\newblock {\em Journal of Cryptology}, 18(2):133--165, Apr 2005.

\bibitem{Rusca2018}
Davide Rusca, Alberto Boaron, Fadri Gr{\"{u}}nenfelder, Anthony Martin, and
  Hugo Zbinden.
\newblock Finite-key analysis for the 1-decoy state qkd protocol.
\newblock {\em Applied Physics Letters}, 112(17):171104, 2018.

\bibitem{Bennett2014_BB84}
Charles~H. Bennett and Gilles Brassard.
\newblock {Quantum cryptography: Public key distribution and coin tossing}.
\newblock {\em Theor. Comput. Sci.}, 560(P1):7--11, dec 2014.

\bibitem{Song2006}
Jian Song, Qi~An, and Shubin Liu.
\newblock A high-resolution time-to-digital converter implemented in
  field-programmable-gate-arrays.
\newblock {\em IEEE Trans. Nuc. Sci.}, 53(1):236--241, 02 2006.

\bibitem{Fishburn2013}
M.~Fishburn, L.~H. Menninga, C.~Favi, and E.~Charbon.
\newblock A 19.6 ps, fpga-based tdc with multiple channels for open source
  applications.
\newblock {\em IEEE Trans. Nuc. Sci.}, 60(3):2203--2208, 06 2013.

\bibitem{Chen2019}
Haochang Chen and David Day-Uei Li.
\newblock Multichannel, low nonlinearity time-to-digital converters based on 20
  and 28 nm fpgas.
\newblock {\em IEEE Transactions on Industrial Electronics}, 66(4):3265--3274,
  2019.

\bibitem{TDC-GPX}
ScioSense.
\newblock Tdc-gpx high-end time-to-digital converter.
\newblock
  \url{https://www.sciosense.com/products/time-to-digital-converters/tdc-gpx-high-end-time-to-digital-converter/}.

\bibitem{Gabriel2010a}
Christian Gabriel, Christoffer Wittmann, Denis Sych, Ruifang Dong, Wolfgang
  Mauerer, Ulrik~L. Andersen, Christoph Marquardt, and Gerd Leuchs.
\newblock {A generator for unique quantum random numbers based on vacuum
  states}.
\newblock {\em Nature Photonics}, 4(10):711--715, oct 2010.

\end{thebibliography}

\end{document}